\documentclass[nopreprintline]{elsarticle_mod}

\usepackage{hyperref}
\usepackage{geometry}                
\usepackage{graphicx}
\usepackage[usenames,dvipsnames]{xcolor}
\usepackage{amssymb}
\usepackage{amsmath}
\usepackage{amsfonts}
\usepackage{epstopdf}
\usepackage{multirow}
\graphicspath{ {./} }

\begin{document}

\title{A Search for Neutrino Emission from the Fermi Bubbles with the ANTARES Telescope}
\author[UPV]{S.~Adri\'an-Mart\'inez}
\author[Colmar]{A.~Albert}
\author[CPPM]{I.~Al~Samarai}
\author[UPC]{M.~Andr\'e}
\author[Erlangen]{G.~Anton}
\author[IRFU/SEDI]{S.~Anvar}
\author[UPV]{M.~Ardid}
\author[NIKHEF]{T.~Astraatmadja\fnref{tag:1}}
\author[CPPM]{J-J.~Aubert}
\author[APC]{B.~Baret}
% ...
\newpage

\address[UPV]{\scriptsize{Institut d'Investigaci\'o per a la Gesti\'o Integrada de les Zones Costaneres (IGIC) - Universitat Polit\`ecnica de Val\`encia. C/ Paranimf 1, 46730 Gandia, Spain}}
\address[Colmar]{\scriptsize{GRPHE - Institut universitaire de technologie de Colmar, 34 rue du Grillenbreit BP 50568 - 68008 Colmar, France}}
\address[CPPM]{\scriptsize{CPPM, Aix-Marseille Universit\'e, CNRS/IN2P3, Marseille, France}}
\address[UPC]{\scriptsize{Technical University of Catalonia, Laboratory of Applied Bioacoustics, Rambla Exposici\'o, 08800 Vilanova i la Geltr\'u,Barcelona, Spain}}
% ...
\address[Erlangen]{\scriptsize{Friedrich-Alexander-Universit\"at Erlangen-N\"urnberg, Erlangen Centre for Astroparticle Physics, Erwin-Rommel-Str. 1, 91058 Erlangen, Germany}}
\address[IRFU/SEDI]{\scriptsize{Direction des Sciences de la Mati\`ere - Institut de recherche sur les lois fondamentales de l'Univers - Service d'Electronique des D\'etecteurs et d'Informatique, CEA Saclay, 91191 Gif-sur-Yvette Cedex, France}}
\address[NIKHEF]{\scriptsize{Nikhef, Science Park, Amsterdam, The Netherlands}}
\address[APC]{\scriptsize{APC, Universit\'e Paris Diderot, CNRS/IN2P3, CEA/IRFU, Observatoire de Paris, Sorbonne Paris Cit\'e, 75205 Paris, France}}
% ...
\address[IRFU/SPP]{\scriptsize{Direction des Sciences de la Mati\`ere - Institut de recherche sur les lois fondamentales de l'Univers - Service de Physique des Particules, CEA Saclay, 91191 Gif-sur-Yvette Cedex, France}}
% ...

\fntext[tag:1]{\scriptsize{Also at University of Leiden, the Netherlands}}
\fntext[tag:2]{\scriptsize{On leave of absence at the Humboldt-Universit\"at zu Berlin}}
\fntext[tag:3]{\scriptsize{Also at Accademia Navale de Livorno, Livorno, Italy}}
\fntext[tag:4]{\scriptsize{Now at the Max Planck Institute for Physics, Munich}}
\fntext[tag:5]{\scriptsize{Now at Academia Sinica, 128 Academia Road, Section 2, Nankang, Taipei 115, Taiwan (R.O.C.) and National Central University, No.300, Jhongda Rd., Jhongli City, Taoyuan County 32001, Taiwan (R.O.C.)}}
\cortext[ca]{Corresponding author \href{mailto:vladimir.kulikovskiy@ge.infn.it}{vladimir.kulikovskiy@ge.infn.it}}

\begin{abstract}
Analysis of the Fermi-LAT data has revealed two extended structures above and below the Galactic Centre emitting gamma rays with a hard spectrum, the so-called Fermi bubbles. 
Hadronic models attempting to explain the origin of the Fermi bubbles predict the emission of high-energy neutrinos and gamma rays with similar fluxes. The ANTARES detector, a neutrino telescope located in the Med\-i\-ter\-ranean Sea, has a good visibility to the Fermi bubble regions.
Using data collected from 2008 to 2011 no statistically significant excess of events is observed and therefore upper limits on the neutrino flux in TeV range from the Fermi bubbles are derived for various assumed energy cutoffs of the source.
\end{abstract}
\maketitle
\section{Introduction}
\begin{figure}
\begin{minipage}{0.5\textwidth}
\center
\includegraphics[width=0.4\textwidth]{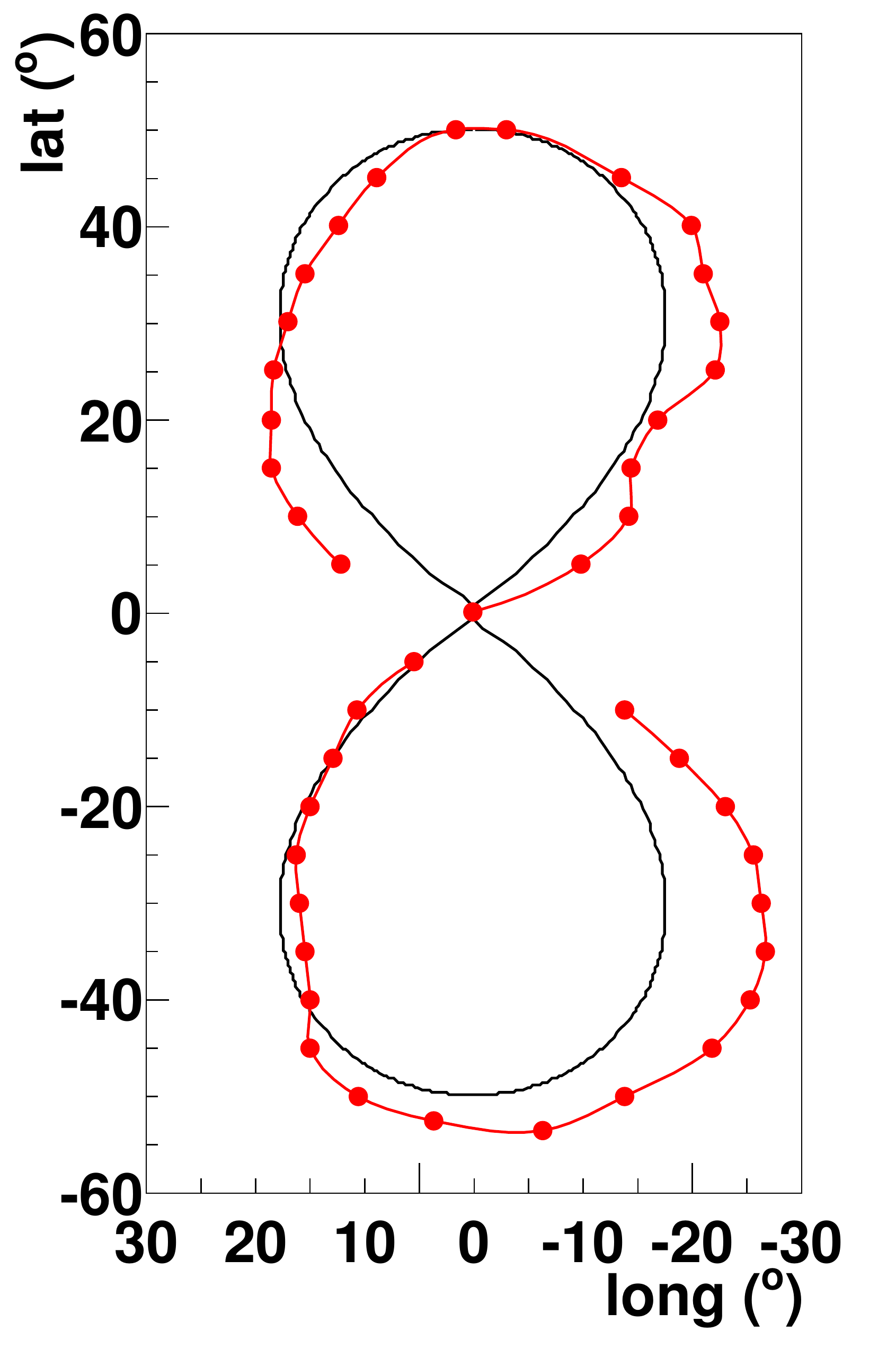}
\end{minipage}
\begin{minipage}{0.5\textwidth}
\caption{Approximate edges (red line, circles) of the north and south Fermi bubbles respectively in galactic coordinates identified from the 1--5 GeV maps built from the Fermi-LAT data~\citep{Su}. The contour line is discontinuous at the region of the Galactic Centre as the maps are severely compromised by the poor subtraction and interpolation over a large number of point sources in this region. The simplified shape of the Fermi bubbles used in this analysis (black line) has an angular area of 0.66 sr.}
\label{fig:fb_shape}
\end{minipage}
\end{figure}

Analysis of data collected by the Fermi-LAT experiment has revealed two large circular structures near the Galactic Centre, above and below the galactic plane~--- the so-called Fermi bubbles~\citep{Su}. The approximate edges of the Fermi bubble regions are shown in Figure~\ref{fig:fb_shape}. These structures are characterised by gamma-ray emission with a hard $E^{-2}$ spectrum and a constant intensity over the whole emission region.

Signals from roughly the Fermi bubble regions were also observed in the microwave band by WMAP~\citep{Dobler} and, recently, in the radio-wave band~\citep{radio}. Moreover, the edges correlate with the X-ray emission measured by ROSAT~\citep{ROSAT}. Several proposed models explaining the emission include hadronic mechanisms, in which gamma rays together with neutrinos are produced by the collisions of cosmic-ray protons with interstellar matter~\citep{CrockerAharonian,Lacki,Thoudam}. Others which include leptonic mechanisms or dark matter decay would produce lower neutrino emission or none at all~\citep{Su,Lacki,Chernyshov,MertschSarkar,dm}. The observation of a neutrino signal from the Fer\-mi bubble regions would play a unique role in discriminating between models.

The properties of the hypothesised neutrino e\-mis\-sion are described in Section~\ref{s:fermi}. An overview of the ANTARES neutrino detector is given in Section~\ref{s:Antares} and the neutrino event reconstruction is described in Section~\ref{s:reconstruction}. The search for neutrino emission is performed by comparing the number of events in the Fermi bubble regions to the number found in similar off-zone regions (Section~\ref{s:offzones}). The event selection optimisation is based on a simulation of the expected signal as described in Section~\ref{s:simulation}. The selected events are presented in Section~\ref{s:results} together with the significance and the upper limit on the neutrino flux from the Fermi bubbles.
\section{Estimation of the neutrino flux}
\label{s:fermi}
The estimated photon flux in the energy range 1--100~GeV covered by the Fermi-LAT detector from the Fermi bubble regions is~\citep{Su}:
\begin{equation}
E^{2}\frac{\mathrm{d}\Phi_\gamma}{\mathrm{d}E} \approx 3-6\times10^{-7}\mathrm{\,GeV\,cm^{-2}\,s^{-1} sr^{-1}}.
\label{f:gamma}
\end{equation}
Assuming a hadronic model in which the gamma-ray and neutrino fluxes arise from the decay of neutral and charged pions respectively, the $\nu_\mu$ and $\overline\nu_\mu$ fluxes are proportional to the gamma-ray flux with proportionality coefficients of 0.211 and 0.195 respectively~\citep{VillanteVissani}.
With this assumption and using~(\ref{f:gamma}) the expected neutrino flux is:
\begin{equation}
E^{2}\frac{\mathrm{d}\Phi_{\nu_\mu+\overline\nu_\mu}}{\mathrm{d}E} = A_\mathrm{theory},
\label{f:fb_flux}
\end{equation}
\begin{equation}
A_\mathrm{theory} \approx 1.2-2.4\times10^{-7}\mathrm{\,GeV\,cm^{-2}\,s^{-1} sr^{-1}}.
\label{f:a_flux}
\end{equation}
The neutrino flux, as well as the gamma-ray flux, is expected to have an exponential energy cutoff, so the extrapolation of~(\ref{f:fb_flux}) towards higher energies can be represented by: 
\begin{equation}
E^{2}\frac{\mathrm{d}\Phi_{\nu_\mu+\overline\nu_\mu}}{\mathrm{d}E} = A_\mathrm{theory}\mathrm{e}^{-E/E^\mathrm{cutoff}_\nu}.
\label{f:fb_flux2}
\end{equation}
The cutoff is determined by the primary protons which have a suggested cutoff $E^\mathrm{cutoff}_p$ in the range from 1~PeV to 10~PeV \citep{CrockerAharonian}. The corresponding neutrino-energy cutoff may be estimated by assuming that the energy transferred from $p$ to $\nu$ derives from the fraction of energy going into charged pions ($\sim20\%$) which is then distributed over four leptons in the pion decay. Thus: 
\begin{equation}
E^\mathrm{cutoff}_\nu \approx E^\mathrm{cutoff}_p/20,
\label{f:cutoff}	
\end{equation}
which gives a range from 50~TeV to 500~TeV for $E^\mathrm{cutoff}_\nu$.

\section{The ANTARES neutrino telescope}
\label{s:Antares}
The ANTARES telescope is a deep-sea Cherenkov detector which is located 40~km from Toulon (France), at a latitude of $42^\circ48'$~N and a mooring depth of 2475 m. at a mooring depth of 2475~m. The energy and direction of incident neutrinos are measured by detecting the Cherenkov light produced in water from muons originating in the charged-current interactions of $\nu_{\mu}$ and $\bar{\nu}_{\mu}$. The light is detected with a three-dimensional array of twelve detection lines comprising 885 optical modules, each containing a 10~inch PMT. More details on the detector construction, its positioning system and the time calibration can be found in~\citep{ANTARES,Positioning,timing}.

The ANTARES detector started data-taking with the first 5 lines installed in 2007. The construction of the detector was completed, with installation of the last two lines, in May 2008. The apparatus has been operating continuously ever since. Its main goal is the detection of neutrinos produced by the cosmic sources.  Muons and neutrinos created in cosmic-ray induced atmospheric showers provide the two main background components for the search for cosmic neutrinos. Although the more than 2 km of water above the detector acts as a partial shield against the atmospheric muons, the downgoing atmospheric muon background at these depths is still bigger than the expected signal. Therefore, the search for cosmic signal concentrates on upgoing events which corresponds to neutrinos which have crossed the Earth. Also, the optical modules are oriented downwards at $45^\circ$ to favour the detection of upgoing particles. The ANTARES neutrino telescope has an excellent visibility by means of the upgoing neutrinos to the Galactic Centre region and to the Fermi bubbles. Since atmospheric neutrinos may traverse the Earth and lead to upgoing tracks in the detector, any signal from the Fermi bubbles would be inferred by observing a significant statistical excess over the background. The signal-to-noise ratio can be improved by rejecting low-energy neutrino events, as the spectrum of the atmospheric neutrinos is steeper than the expected source spectrum.

\section{Track and energy reconstruction}
\label{s:reconstruction}
The track of a muon passing through the detector is reconstructed using the arrival time of the photons together with the positions and orientations of the photomultipliers. Details of the tracking algorithm are given in~\citep{pointsearch}. Only events  reconstructed as upgoing have been selected. In addition, cuts on the reconstruction quality parameters have been applied in order to reject downgoing atmospheric muon events that are incorrectly reconstructed as upgoing tracks. These parameters are the quality $\Lambda$ of the track fit, which is derived from the track fit likelihood, and the uncertainty $\beta$ of the reconstructed track direction. The choice of the cut on $\Lambda$ fixes the amount of background from misreconstructed atmospheric muons in the neutrino sample. Neutrino simulations for an $E^{-2}$ neutrino spectrum have yielded a median angular resolution on the neutrino direction of less than $0.6^\circ$ for events with $\Lambda > -5.2$ and $\beta<1^\mathrm{o}$~\citep{pointsearch}.

Shower-like events are identified by using a second tracking algorithm with $\chi^2$-like fit, assuming the hypothesis of a relativistic muon ($\chi^2_\mathrm{track}$) and that of a shower-like event ($\chi^2_\mathrm{point}$)~\citep{bbfit}. Events with better point-like fit ($\chi^2_\mathrm{point}<\chi^2_\mathrm{track}$) have been excluded from the analysis.

In this analysis the energy of the muons entered or born in the detector was estimated using Artificial Neural Networks, which are produced using a machine learning algorithm which derives the dependence between a set of observables and the energy estimate in a semi-parametric way~\citep{Jutta}. The parameters used include the number of detected photons, and the total deposited charge. The median resolution for $\log_{10}E_\mathrm{Rec}$ is about 0.3 for muons with an energy of 10~TeV. The reconstructed energy $E_\mathrm{Rec}$ is used to reject the atmospheric neutrino background while $\Lambda$ is used mostly to reject atmospheric muons. The choice of cuts on $\Lambda$ and $E_\mathrm{Rec}$ in this work is discussed in Section~\ref{s:MRF}.

\section{Off-zones for background estimation}
\label{s:offzones}
A signal from the combined Fermi bubble regions is searched for by comparing the number of selected events from the area of both bubbles (on-zone) to that of similar regions with no expected signal (off-zones). The simplified shape of each Fermi bubble as used in this analysis is shown in Figure~\ref{fig:fb_shape}.

\begin{figure}
\center
\includegraphics[width=0.8\textwidth]{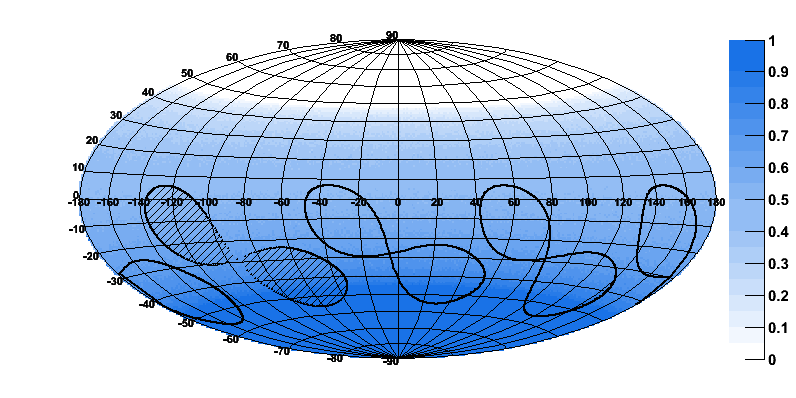}
\caption{Hammer equal-area map projection in equatorial coordinates $(\alpha,\delta)$ showing the Fermi bubble regions (on-zone) shaded area in the centre. The regions corresponding to the three off-zones are also depicted. The colour fill represents the visibility of the sky at the ANTARES site. The maximum on the colour scale corresponds to a 24~h per day visibility.}
\label{fig:fb_offzones_eq}
\end{figure}

Off-zones are defined as fixed regions in equatorial coordinates which have identical size and shape as the on-zone but have no overlap with it. In local coordinates, such off-zones have the same, sidereal-day periodicity as the on-zone and span the same fraction of the sky, but with some fixed delay in time. The size of the Fermi bubbles allows at maximum three non-overlapping off-zones to be selected. The on-zone and three off-zones are shown in Figure~\ref{fig:fb_offzones_eq} together with the sky visibility. The visibility of each point on the sky is the fraction of the sidereal day during which it is below the horizon at the ANTARES site (in order to produce upgoing events in the detector). The average visibility of the Fermi bubbles is 0.68 (0.57 for the northern bubble and 0.80 for the southern bubble) and it is the same for the off-zones.

Slightly changing detector efficiency with time and gaps in the data acquisition can produce differences in the number of background events between the on-zone and the three off-zones. In order to test for such an effect, firstly, the number of events in the off-zones is extracted from the data for various cuts ($\Lambda^\mathrm{cut}$, $E_\mathrm{Rec}^\mathrm{cut}$) and the difference in the event numbers between each pair of off-zones is calculated. This difference is compared with the statistical uncertainty and no excess is seen beyond the expected statistical fluctuations. Secondly, the number of events in the on-zone together with the average number of events in the three off-zones is tested using the simulated atmospheric background and the difference is found to be within the expectation from the statistical uncertainty. It can be concluded, therefore, that this effect is negligible.
\section{Event selection criteria}
\label{s:MRF}
The analysis adopts a blind strategy in which the cut optimisation is performed using simulated data for the signal and the background. The main quantities used to discriminate between the cosmic neutrino candidate events and the background from misreconstructed atmospheric muons and from atmospheric neutrinos are the tracking quality parameter $\Lambda$ and the reconstructed muon energy $E_\mathrm{Rec}$. 

\label{s:simulation}
The simulation chain for ANTARES is described in~\citep{Brunner}. For the expected signal from the Fermi bubbles, the $\nu_\mu$ and $\overline\nu_\mu$ fluxes according to Section~\ref{s:fermi} are assumed, using four different cutoffs $E^\mathrm{cutoff}_\nu$: no cutoff ($E^\mathrm{cutoff}_\nu=\infty$), 500~TeV, 100~TeV and 50~TeV. Atmospheric neutrinos are simulated using the model from the Bartol group~\citep{Bartol} which does not include the decay of charmed particles.

\begin{figure}
\center
\includegraphics[width=0.7\textwidth]{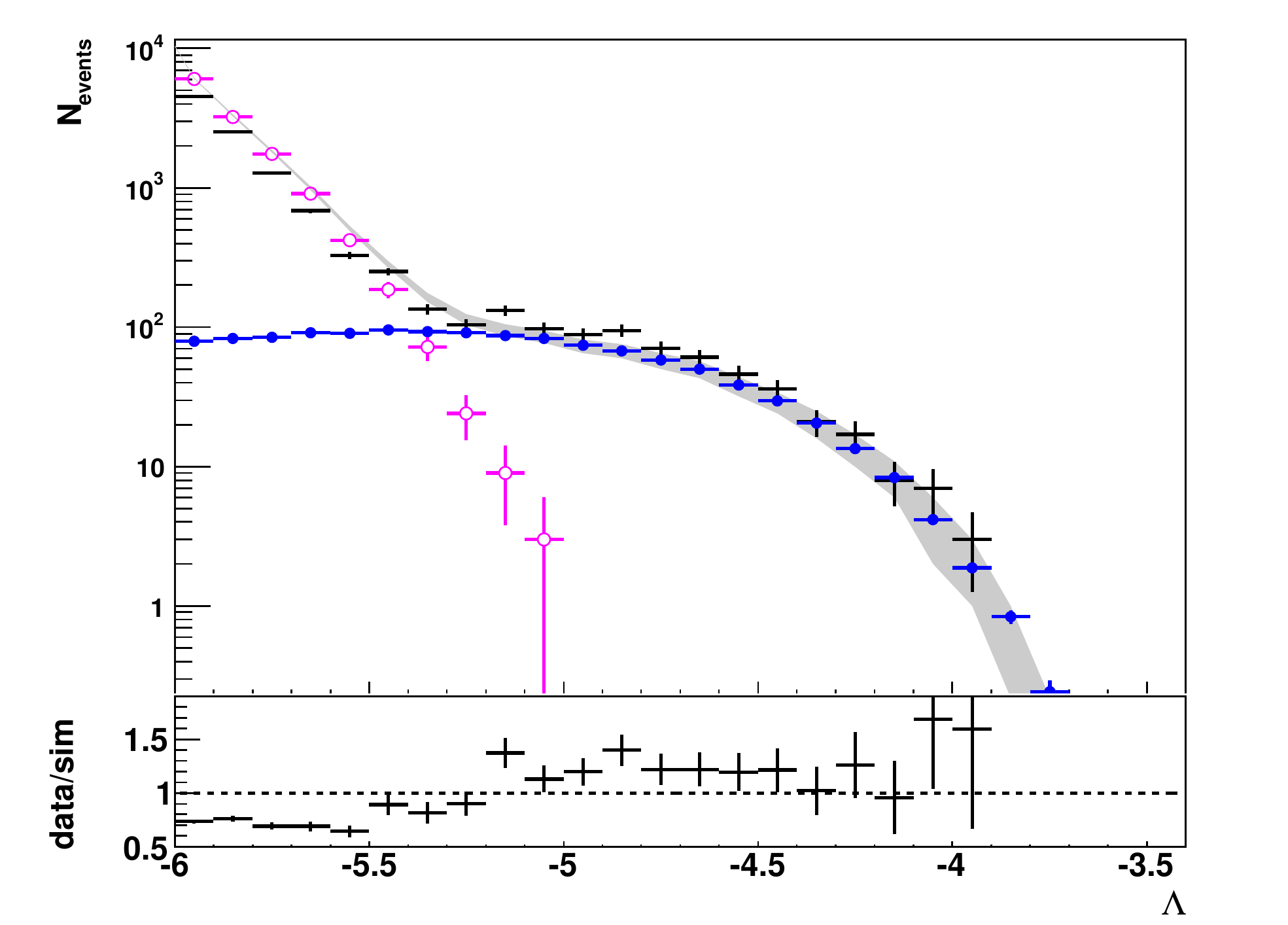}
\caption{Distribution of the fit-quality parameter $\Lambda$ for the upgoing events arriving from the three off-zones: data (black crosses), 68\% confidence area given by the total background simulation (grey area), $\nu^{\mathrm{sim}}_{\mathrm{atm}}$ (blue filled circles), $\mu^{\mathrm{sim}}_{\mathrm{atm}}$ (pink empty circles); bin-ratio of the data to the total background simulation (bottom).}
\label{fig:fb_lambda}
\end{figure}

Data in the period from May 2008, when the detector started to operate in its complete configuration, until December 2011 are used. The total livetime selected for this analysis amounts to 806~days. Figure~\ref{fig:fb_lambda} shows the distribution of data and simulated events as a function of the parameter $\Lambda$ for events arriving from the three off-zones. Here the events with at least 10 detected photons and the angular error estimate $\beta<1^\circ$ are selected. The requirement on the number of photons removes most of the low-energy background events. The angular error condition is necessary in order to ensure a high angular resolution to avoid events originating from an off-zone region being associated with the signal region and vice versa.

At $\Lambda\sim-5.3$ the main background component changes from the misreconstructed atmospheric muons to the upgoing atmospheric neutrino events as seen in Figure~\ref{fig:fb_lambda}. The flux of atmospheric neutrinos in the simulation is 23\% lower than observed in the data. This is well within the systematic uncertainty on the atmospheric neutrino flux and the atmospheric flux from the simulations was scaled accordingly in the following analysis. 

\begin{figure}
\center
\includegraphics[width=0.7\textwidth]{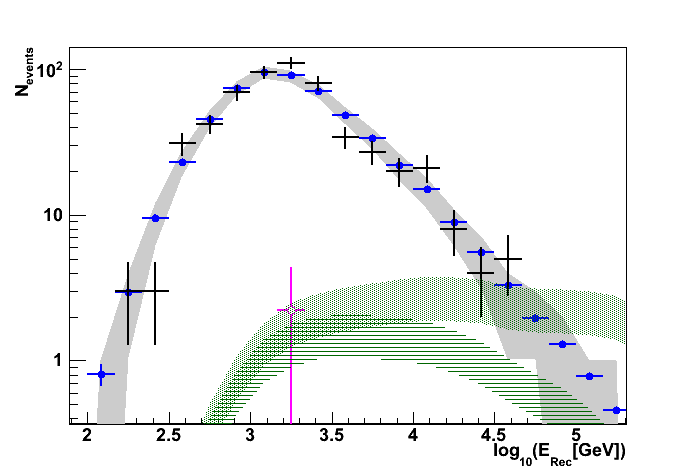}
\caption{$E_{\mathrm{Rec}}$ distribution of the events arriving from the three off-zones with $\Lambda>-5.1$: data (black crosses), 68\% confidence area for the total background from simulation (grey area), $\nu^{\mathrm{sim}}_{\mathrm{atm}}$ (blue filled circles), $\mu_{\mathrm{atm}}^\mathrm{sim}$ (pink empty circles), expected signal from the Fermi bubbles according to~(\ref{f:a_flux}--\ref{f:fb_flux2}) without neutrino energy cutoff (green dotted area) and with 50~TeV energy cutoff (green dashed area). The expected signal was scaled by a factor of 3 to allow easy comparison with the total off-zone distribution.}
\label{fig:fb_eann}
\end{figure}

A comparison of the energy estimator for data and for atmospheric neutrino simulation is shown in Figure~\ref{fig:fb_eann} for the same event selection but with a stricter cut $\Lambda>-5.1$ to remove most of the misreconstructed atmospheric muons. The reconstructed energy of all simulated events has been shifted, $\log_{10}E_\mathrm{Rec} = \log_{10}E_\mathrm{Rec}^\mathrm{original} + 0.1$, in order to improve the agreement between data and simulations. This is within the estimated uncertainty of the optical module efficiency and the water absorption length~\citep[Figure 4.24]{Dimitris}.
 
The final event selection is optimised by minimising the average upper limit on the flux:
\begin{equation}
\overline \Phi_{90\%} = \Phi_{\nu_\mu+\overline\nu_\mu} \frac{\overline s_{90\%}(b)}{s},
\label{eq:sensitivity}
\end{equation}
where $s$ is the number of events simulated with the flux $\Phi_{\nu_\mu+\overline\nu_\mu}$ from~(\ref{f:fb_flux2}). The method uses an approach following Feldman \& Cousins~\cite{FeldmanCousins} to calculate signal upper limits with 90\% confidence level, $\overline s_{90\%}(b)$, for a known number of simulated background events $b$.
This best average upper limit in the case of no discovery represents the sensitivity of the detector to the Fermi bubbles' flux~\citep{MRF}. 
Using~(\ref{f:fb_flux2}) the average upper limit on the flux coefficient $A$ can be defined as:
\begin{equation}
\overline A_{90\%} = A_\mathrm{theory}\frac{\overline s_{90\%}(b)}{s}.
\label{eq:sensitivityA}
\end{equation}

Table~\ref{table:MRF} reports the optimal cuts ($\Lambda^\mathrm{cut}$, $E_\mathrm{Rec}^\mathrm{cut}$) obtained for the four chosen cutoff energies ($\infty$, 500, 100, 50~TeV) of the neutrino source spectrum and the corresponding value of the average upper limit on the flux coefficient $\overline  A_{90\%}$. Additionally, the optimal cuts for $E^\mathrm{cutoff}_\nu = 100$~TeV are applied for the other neutrino-energy cutoffs and the values $\overline A_{90\%}^{100}$ are reported for comparison. As the obtained values $\overline A_{90\%}$ and $\overline A_{90\%}^{100}$ for each cutoff are similar, the 100~TeV cuts are chosen for the final event selection.
\begin{table}
\caption{Optimisation results for each cutoff of the neutrino energy spectrum. Average upper limits on the flux coefficient $\overline A_{90\%}$ are presented in units of $10^{-7} \mathrm{\,GeV\,cm^{-2}\,s^{-1} sr^{-1}}$. Numbers with a star indicate the cut used for the $\overline A^{100}_{90\%}$ calculation presented in the last row of the table.}
\centering
\label{table:MRF}
\begin{tabular}{l r r r r}
$E_\nu^\mathrm{cutoff}$ (TeV)&$\infty$&500&$100\;\;$&50\\
\hline 
$\Lambda^\mathrm{cut}$&$-5.16$&$-5.14$&$-5.14^{*}$&$-5.14$\\
$\log_{10}(E_\mathrm{Rec}^\mathrm{cut}[\mathrm{GeV}])$&4.57&4.27&$4.03^{*}$&3.87\\
$\overline A_{90\%}$&2.67&4.47&$8.44\;\,$&12.43\\
$\overline A_{90\%}^\mathrm{100}$ (100 TeV cuts)&3.07&4.68&$8.44\;\,$&12.75\\
\end{tabular}
\end{table}

At energies above 100~TeV the semi-leptonic decay of short-lived charmed particles might become a major source of atmospheric neutrino background. The uncertainty in the flux from this contribution is large~\citep{Costa, Martin, Sarcevic}. Due to the comparison of on and off zones (Section~\ref{s:offzones}) and the final cut $\sim10$~TeV (Table~\ref{table:MRF}) the flux from charmed particle decays will not have a significant impact on the analysis nor alter the final result on upper limits.

\section{Results}
\label{s:results}
The final event selection $\Lambda>-5.14$, $\log_{10}(E_\mathrm{Rec}[\mathrm{GeV}])>4.03$  is applied to the unblinded data. In the three off-zones 9, 12 and 12 events are observed. In the Fermi bubble regions $N_\mathrm{obs}=16$ events are measured. This corresponds to 1.2~$\sigma$ excess calculated using the method by Li \& Ma~\cite{LiMa}.

\begin{figure}
\center
\includegraphics[width=0.7\textwidth]{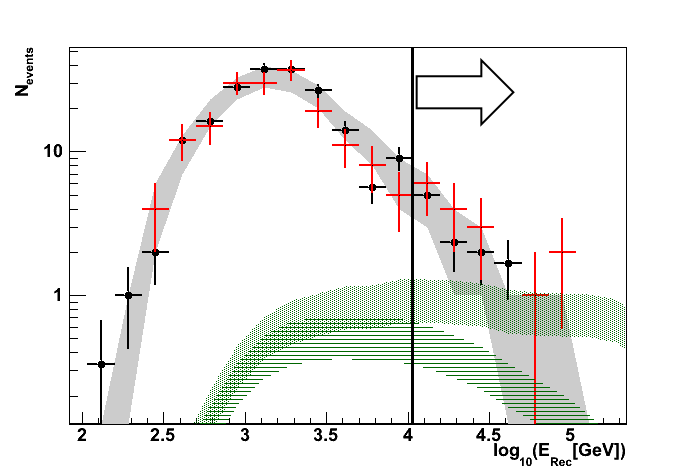}
\caption{Distribution of the reconstructed energy of the events after the final cut on $\Lambda$: events in on-zone (red crosses), average over off-zones (black circles), 68\% confidence area given by the total background simulation (grey area), expected signal from the Fermi bubbles without neutrino-energy cutoff (green dotted area) and 50~TeV cutoff (green dashed area). The chosen $E^\mathrm{cut}_\mathrm{Rec}$ is represented by the black line with an arrow.}
\label{fig:unblinded_events}
\end{figure}

The distribution of the energy estimator for both the on-zone and the average of the off-zones is presented in Figure~\ref{fig:unblinded_events}. A small excess of high-energy events in the on-zone is seen with respect to both the average from the off-zones and the atmospheric neutrino simulation.

An upper limit on the number of signal events is calculated using a Bayesian approach at 90\% coverage using the probability distribution with two Poisson distributions for the measurements in the on-zone and in the three off-zones. In order to account for systematic uncertainties in the simulation of the signal, a dedicated study has been performed in which the assumed absorption length in seawater is varied by $\pm10\%$ and the assumed optical module efficiency is varied by $\pm10\%$. For each variation the number of events is calculated for each cutoff and compared with the number of signal events $s$ obtained using the standard simulation. The differences are calculated and summed in quadrature to obtain $\sigma_\mathrm{syst}$. A Gaussian distribution of the efficiency coefficient for the signal with mean $s$ and standard deviation $\sigma_\mathrm{syst}$ is convoluted to the probability distribution. The maximum of the probability distribution is found for every neutrino flux coefficient $A$ and the obtained profile likelihood is used together with the flat prior for $A$ to calculate the post-probability. The upper and lower limits for $A$ are extracted from the post-probability to have 90\% coverage.

\begin{figure}
\center
\includegraphics[width=0.7\textwidth]{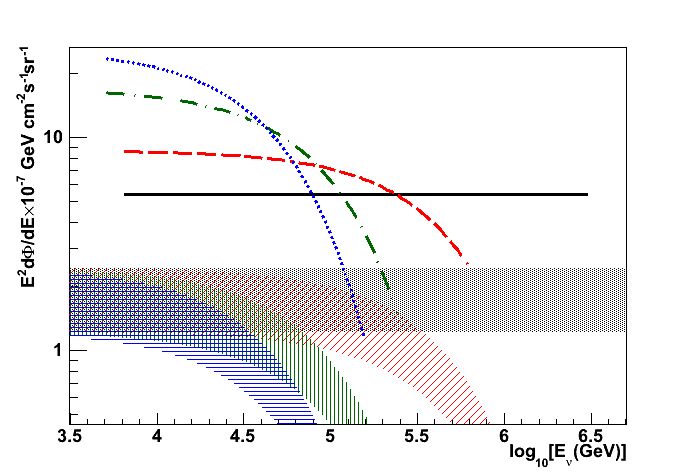}
\caption{Upper limits on the neutrino flux from the Fermi bubbles for different cutoffs: no cutoff (black solid), 500~TeV (red dashed), 100~TeV (green dot-dashed), 50~TeV (blue dotted) together with the theoretical predictions for the case of a purely hadronic model (the same colours, areas filled with dots, inclined lines, vertical lines and horizontal lines respectively). The limits are drawn for the energy range where 90\% of the signal is expected.}
\label{fig:upper_limit_fluxes}
\end{figure}

The results are summarised in Table~\ref{table:flux}. A graphical representation of the upper limits on a possible neutrino flux together with the predicted flux is shown in Figure~\ref{fig:upper_limit_fluxes}. The obtained upper limits are above the expectations from the considered models. The modified Feldman \& Cousins approach with the included uncertainties gives comparable results~\citep{Conrad}. 
\begin{table}
\center
\caption{90\% confidence level upper limits on the neutrino flux coefficient $A_{90\%}$ for the Fermi bubbles presented in units of $10^{-7} \mathrm{\,GeV\,cm^{-2}\,s^{-1} sr^{-1}}$.}
\label{table:flux}
\begin{tabular}{r c c c c}
$E_\nu^\mathrm{cutoff}$ (TeV)  & $\infty$  & 500 & 100 & 50 \\
\hline
number of signal events in simulation $s$ & 2.9 & 1.9 & 1.1 & 0.7 \\
uncertainty on the efficiency $\sigma_\mathrm{syst}$, \% & 14 & 19 & 24 & 27 \\
$A_{90\%}$ & 5.4 & 8.7 & 17.0 & 25.9 \\
\end{tabular}
\end{table}

\section{Conclusions}
High-energy neutrino emission from the region of the Fermi bubbles has been searched for using data from the ANTARES detector. An analysis of the 2008--2011 ANTARES data yield\-ed a 1.2 $\sigma$ excess of events in the Fermi bubble regions, compatible with the no-signal hypothesis. For the optimistic case of no energy cutoff in the flux, the upper limit is within a factor of three of a prediction from the purely hadronic model based on the measured gamma-ray flux. The sensitivity will improve as more data is accumulated (more than 65\% gain in the sensitivity is expected once 2012--2016 data is added to the analysis). The next generation KM3NeT neutrino telescope will provide more than an order of magnitude improvement in sensitivity~\citep{km3,km3theoretical1,km3theoretical2}.

\section{Acknowledgements}
The authors acknowledge the financial support of the funding agencies: Centre National de la Recherche Scientifique (CNRS), Commissariat \'a l'\'Energie Atomique et aux \'Energies Alternatives (CEA), Agence National de la Recherche (ANR), Commission Europ\'eenne (FEDER fund and Marie Curie Program), R\'egion Alsace (contrat CPER), R\'egion Provence-Alpes-C\^ote d'Azur, D\'epartement du Var and Ville de La Seyne-sur-Mer, France; Bundesministerium f\"ur Bildung und Forschung (BMBF), Germany; Istituto Nazionale di Fisica Nucleare (INFN), Italy; Ministerio de Ciencia e Innovaci\'on (MICINN), Prometeo of Generalitat Valenciana and MultiDark, Spain; Agence de l'Oriental, Morocco; Stichting voor Fundamenteel Onderzoek der Materie (FOM), Nederlandse organisatie voor Wetenschappelijk Onderzoek (NWO), the Netherlands; National Authority for Scientific Research (ANCS-UEFISCDI), Romania; Council of the President of the Russian Federation for young scientists and leading scientific schools supporting grants, Russia. Technical support of Ifremer, AIM and Foselev Marine for the sea operation and the CC-IN2P3 for the computing facilities is acknowledged.

\end{document}